\documentclass[a4paper]{jpconf}
\usepackage{graphicx}
\usepackage{dcolumn}
\usepackage{bm}
\usepackage{pstricks}
\usepackage{amssymb,amsmath}

\newcommand{\ba}{\begin{eqnarray}}
\newcommand{\ea}{\end{eqnarray}}
\def\be{\begin{equation}}
\def\ee{\end{equation}}

\newcommand{\op}[1]{ {\bf #1}}

\def\prl#1#2#3{Phys.\ Rev.\ Lett.\ {\bf #1}, #2 (#3)}
\def\prb#1#2#3{Phys.\ Rev.\ B {\bf #1}, #2 (#3)}

\def\bimno{Bi$_3$Mn$_4$O$_{12}$(NO$_3$)}

\begin{document}

\title{ Self consistent study of the quantum phases in a 
frustrated antiferromagnet on the bilayer honeycomb lattice.}

\author{ M.\ Arlego}
\address{IFLP - CONICET, Departamento de F\'isica, Universidad Nacional de La Plata,
C.C.\ 67, 1900 La Plata, Argentina.}

\author{ C. A.\ Lamas}
\address{IFLP - CONICET, Departamento de F\'isica, Universidad Nacional de La Plata,
C.C.\ 67, 1900 La Plata, Argentina.}

\author{Hao Zhang}
\address{Institute for Solid State Physics, University of Tokyo, Kashiwa, Chiba 277-8581, Japan}

\begin{abstract}
We study the frustrated Heisenberg model on the bilayer
honeycomb lattice. The ground-state energy and spin gap are calculated, using
different bosonic representations at mean field level and numerical calculations,
to explore different sectors of the phase diagram.
In particular we make use of a bond operator formalism and series expansion calculations
to study the extent of dimer inter-layer phase.
On the other hand we use the Schwinger boson method and exact diagonalization on small
systems to analyze the evolution of on-layer phases. In this case we specifically
observe a phase that presents a spin gap and short range N\'eel correlations that
survives even in the presence of non-zero next-nearest-neighbor interaction and inter-layer
coupling.
\end{abstract}


\section{Introduction}

The study of the possible disordered ground states on two-dimensional antiferromagnets has received a
great interest in the last years.
In particular,  the existence of quantum disordered phases has been studied in the phase diagram
of antiferromagnets in a single layer honeycomb lattice
\cite{Cabra_honeycomb_prb,Cabra_honeycomb_2,Zhang_PRB_2013,
Albuquerque,
Li_2012_honeyJ1-J2-J3,QDM_letter,QDM_lamas,Mezzacapo,Ganesh_PRL_2013,
Albuquerque_capponi_2012}. However, the study of the influence of possible interlayer coupling on these phases is scarce \cite{Ganesh_2011,Ganesh_QMC,Oitmaa_2012,Zhang_bilayer}.
From the experimental side, a significative progress on the study of the
bismuth oxynitrate, {\bimno}, has been made by Smirnova \textit{et al.}\cite{BiMnO}. In this material
the Mn$^{4+}$ ions form a honeycomb lattice and two layers of such honeycomb lattices are separated
by bismuth atoms, forming a bilayer structure.
The study of the magnetic susceptibility indicates two-dimensional magnetism
and no long-range ordering down to 0.4 K,
suggesting a nonmagnetic ground state\cite{BiMnO,ESR}.
In addition, density functional studies indicate that dominant interactions are the
interlayer interaction $J_{\bot}$ and the nearest-neighbor interaction $J_1$
on each layer\cite{Kandpal}.
\begin{figure}[t!]
\begin{center}
\includegraphics[width=0.6\columnwidth]{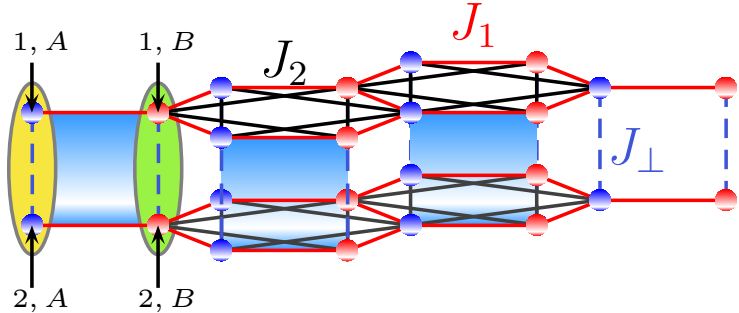}
\caption{(Color online)  Schematic representation of the coupling interactions in \bimno.
Colored areas correspond to the unit cells. The sites in each unit cell are labeled as ($1,A$), ($2,A$), ($1,B$)
and ($2,B$).}
\label{fig:layers}
\end{center}
\end{figure}

The aim of this paper is to study the zero temperature ground state
of the frustrated spin-$1/2$ Heisenberg model on the bilayer honeycomb lattice.
We study the $S=1/2$ case where the quantum fluctuations becomes more important
in order to characterize the quantum phases in the model.
On the other hand, although the material \bimno\ has $S=3/2$, the substitution of
Mn$^{4+}$ in \bimno\ by V$^{4+}$ may lead to the realization of the
$S=1/2$ Heisenberg model on the honeycomb lattice.
We use two different mean field self-consistent approaches based on
bosonic representations of the spin operators to study this system, combined
with Lanczos and series expansion methods to support the mean field results.


\section{Self consistent calculations on the bilayer Model}
We study the following Heisenberg model on the bilayer honeycomb lattice
\ba
 \label{eq:Hspin_general}
 H &=&\sum_{\vec{r},\vec{r}',\alpha,\beta}
 J_{\alpha,\beta}(\vec{r},\vec{r}')
 \vec{\bf{S}}_{\alpha}(\vec{r})\cdot \vec{\bf{S}}_{\beta}(\vec{r}')
\ea
where, $\vec{\bf{S}}_{\alpha}(\vec{r})$ is the spin operator on site $\alpha$
corresponding to the unit cell $\vec{r}$. $\alpha$ takes
the values $\alpha=(1,A),\; (2,A),\; (1,B),\; (2,B)$ corresponding to the four sites on each
unit cell as depicted in Fig. \ref{fig:layers}, together with the
couplings $J_{\alpha,\beta}(\vec{r},\vec{r}')$.
The coupling $J_{\bot}$ does not introduce frustration in the
system and then, at the classical level and $T=0$, it does not affect
the classical N\'eel phase, present for $J_{2}/J_{1}<1/6$.
In the quantum case the situation is much subtle, increasing $J_{\bot}$ N\'eel order is likely
to melt giving rise to a non-magnetic phase.
%
%

For large values of $J_{\bot}$ we expect the ground state to be an
interlayer valence bond crystal (IVBC) with corresponding spins from
both layers forming dimers (as ilustrated in Figure \ref{fig:layers}). This limit represents an  excellent starting point
for the bond operators formalism and series expansion calculations.

On the other hand, starting from the magnetically ordered phase, the N\'eel order can be destroyed both
by increasing the frustration on each layer or
increasing the coupling between layers.
The destruction of N\'eel order in a frustrated single layer honeycomb lattice
has been studied by means of various approaches \cite{Zhang_PRB_2013,Albuquerque,Mezzacapo,Ganesh_2011,Zhang_bilayer,Rastelli,Fouet,Einarsson,Mattsson_PRB_1994}. In the following
we use two different bosonic representations of the spin operators to study
the influence of the interlayer coupling in the ground state of the bilayer model.

\subsection{Bond operators Mean field approach.}

First, we use the well known bond-operator method to study the bilayer antiferromagnet
described by Hamiltonian (\ref{eq:Hspin_general}).
We start by introducing a bond-operator representation of spin operators
\small
\ba
\vec{\bf{S}}^{\alpha}_{\eta,A}(\vec{r})&=&\frac12\left(
(-1)^{\eta+1}\left[
{\bf s}^{\dagger}_{A}(\vec{r}){\bf a}_{\alpha}(\vec{r})+{\bf a}^{\dagger}_{\alpha}(\vec{r}){\bf s}_{A}(\vec{r})\right]
-i \epsilon_{\alpha\beta\gamma} {\bf a}^{\dagger}_{\beta}(\vec{r}){\bf a}_{\gamma}(\vec{r})
\right)\\
\vec{\bf{S}}^{\alpha}_{\eta,B}(\vec{r})&=&\frac12\left(
(-1)^{\eta+1}\left[
{\bf s}^{\dagger}_{B}(\vec{r}){\bf b}_{\alpha}(\vec{r})+{\bf b}^{\dagger}_{\alpha}(\vec{r}){\bf s}_{B}(\vec{r})\right]
-i \epsilon_{\alpha\beta\gamma} {\bf b}^{\dagger}_{\beta}(\vec{r}){\bf b}_{\gamma}(\vec{r})
\right),
\ea
\normalsize
where $\eta=1,2$ is the layer index, $\vec{\bf{S}}^{\alpha}_{\eta,A}(\vec{r})$ is the spin operator
on the sublattice $A$ of  layer $\eta$ corresponding to the unit cell located at $\vec{r}$
(See figure \ref{fig:layers}).
Operators ${\bf s^{\dagger}_{A}}(\vec{r})$ and ${\bf a}^{\dagger}_{\alpha}(\vec{r})$ create singlet and triplets states (out of a vacuum $|0\rangle$) in the vertical bond placed in sub-lattice $A$ and are defined as:
${\bf s}_{A}^{\dagger} |0\rangle = \frac{1}{\sqrt{2}}(|\uparrow \downarrow \rangle -|\downarrow \uparrow \rangle)$,
${\bf a}_{x}^{\dagger} |0\rangle = -\frac{1}{\sqrt{2}}(|\uparrow \uparrow \rangle -|\downarrow \downarrow \rangle)$,
${\bf a}_{y}^{\dagger} |0\rangle = \frac{i}{\sqrt{2}}(|\uparrow \uparrow \rangle +| \downarrow \downarrow \rangle)$,
%
%
${\bf a}_{z}^{\dagger} |0\rangle = \frac{1}{\sqrt{2}}(|\uparrow \downarrow \rangle +| \downarrow \uparrow \rangle)$,
%
and similar expressions for sublattice $B$. 
These kind of representations where proposed by Sachdev\cite{Sachdev}
in order to treat quantum phase transitions between Ne\'el and
dimerized phases. Operators belonging to the same unit cell satisfy the bosonic commutation relations
whereas operators belonging to different unit cells commute.
The restriction that the physical states are either singlets or triplets leads to the constraints
in each unit cell, ${\bf s}_{A}^{\dagger}{\bf s}_{A}+\sum_{\alpha}{\bf a}^{\dagger}_{\alpha}{\bf a}_{\alpha}=1$ and ${\bf s}_{B}^{\dagger}{\bf s}_{B}+\sum_{\alpha}{\bf b}^{\dagger}_{\alpha}{\bf b}_{\alpha}=1$.
Introducing the bond-operators representation of the spin operators in (\ref{eq:Hspin_general})
we obtain a bosonic version of the Hamiltonian.
We transform Fourier and retain terms up to second order to write
${\bf H}={\bf H_{\bot}}+{\bf H_{1}}+{\bf H_{\lambda}}$,
where
\ba
{\bf H_{\bot}}&=&-\frac{3}{2} J_{\bot}s^{2}N+\frac{J_{\bot}}{4}\sum_{\vec{k},\alpha}\left\{
{\bf a}_{\alpha}^{\dagger}(\vec{k}){\bf a}_{\alpha}(\vec{k})+{\bf b}_{\alpha}^{\dagger}(\vec{k}){\bf b}_{\alpha}(\vec{k})
\right\},   \\
{\bf H_{1}}&=&\frac{J_{1}}{4}\sum_{\vec{k},\alpha}\left\{
\gamma(\vec{k})\left(
{\bf a}_{\alpha}(\vec{k}) {\bf b}_{\alpha}(-\vec{k}) +  {\bf a}_{\alpha}(\vec{k}) {\bf b}^{\dagger}_{\alpha}(\vec{k})
 +{\bf b}^{\dagger}_{\alpha}(\vec{k}){\bf a}_{\alpha}(\vec{k}) +   {\bf b}^{\dagger}_{\alpha}(\vec{k}){\bf a}^{\dagger}_{\alpha}(-\vec{k})
\right)\right.\\\nonumber
&+&\left.
\gamma(\vec{k})\left(
{\bf b}_{\alpha}(\vec{k}) {\bf a}_{\alpha}(-\vec{k}) +  {\bf a}^{\dagger}_{\alpha}(\vec{k}) {\bf b}_{\alpha}(\vec{k})
 +{\bf a}^{\dagger}_{\alpha}(\vec{k}) {\bf b}^{\dagger}_{\alpha}(-\vec{k}) +   {\bf b}_{\alpha}(\vec{k}){\bf a}^{\dagger}_{\alpha}(\vec{k})
\right)
\right\},\\
{\bf H_{\lambda}}&=&(2s^{2}-5)N\lambda
+ \lambda\sum_{\vec{k},\alpha}\left\{
 {\bf a}^{\dagger}_{\alpha}(\vec{k}) {\bf a}^{\dagger}_{\alpha}(\vec{k})+{\bf b}^{\dagger}_{\alpha}(\vec{k}) {\bf b}^{\dagger}_{\alpha}(\vec{k})
\right\},
\ea
where, we have assumed that condensation of singlets occurs, \emph{i.e.} $\langle\op{s}_{\alpha}(\vec{r})\rangle=s $,
$\gamma(\vec{k})=s^{2}(1+e^{i\vec{k}\cdot\vec{e}_{1}}+e^{i\vec{k}\cdot\vec{e}_{2}})$, $\vec{e}_{1}$ and
$\vec{e}_{2}$ are the primitive vectors on a triangular lattice and $\lambda$ is a Lagrange multiplier
related to the constraint in the number of bosons. Diagonalization by using a Bogoliubov transformation allows us to write the following
expression for the ground state energy
\ba
\frac{E}{N}=(2s^{2}-5)\lambda-\frac{3}{4}J_{\bot}(2s^{2}+1)+\sum_{\alpha}\int \frac{d^2 k}{V}
\left( \omega^{(A)}_{\alpha}(\vec{k})+\omega^{(B)}_{\alpha}(\vec{k})\right),
\ea
where $\omega^{(A)}_{\alpha}(\vec{k})$ and $\omega^{(B)}_{\alpha}(\vec{k})$ are triplet energies.
The parameters $s^{2}$ and $\lambda$ are determined by solving self-consistently the saddle point conditions
$\frac{\partial E}{\partial \lambda}=0$ and $\frac{\partial E}{\partial s^2}=0$
which are used to evaluate the energy and gap of the system and compare with
numerical techniques.

\begin{figure}
\begin{center}
\includegraphics[width=0.7\columnwidth]{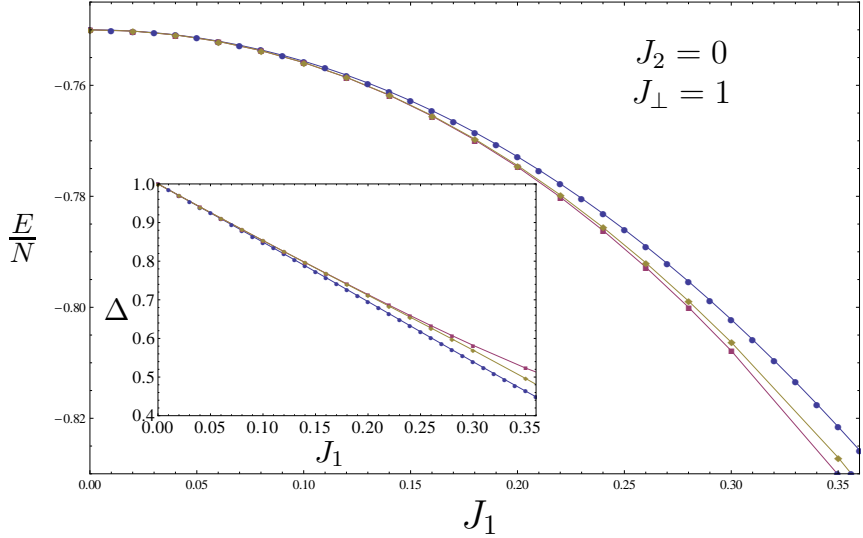}
\caption{(Color online) Ground state energy per dimer as a function of $J_{1}$
obtained by the self consistent bond-operator approach (blue circles),
Lanczos (ED) on a $24$ sites system (red squares) and $O(5)$ Series
Expansion (SE) (yellow rhombi).
Inset:  triplet gap (same set of parameters as main panel)
BO (blue circles) ED (red squares) and SE (yellow rhombi).}
\label{fig:results_BO}
\end{center}
\end{figure}

In order to complement our study, we have performed series expansion
(SE) calculations, starting from the limit of isolated dimers 
connecting spins from both layers via $J_{\bot}$. 
Notice that, this kind of expansions 
remain valid in the same limit that the bond operator approach ({\it i.e.} in the limit of strong
inter-layer coupling.)
To this end we have performed a continuous unitary transformation on the original Hamiltonian, using the flow equation method. This technique allows to obtain perturbatively an effective Hamiltonian that keeps the block diagonal structure of decoupled dimers. We refer for details of the
method to ref.\cite{Knetter2000a}.

For the present model we have performed $O(5)$ and $O(4)$ SE in
$J_{1,2}$ for ground state energy and for triplet dispersion, respectively. Explicit expressions are too long to be printed explicitly but are available electronically upon request.
In Fig.\ref{fig:results_BO} we show the ground state energy per site as a function of $J_1$ and
$J_2=0$, obtained by BO (blue circles), O(5) SE (yellow rhombi) and ED (red squares) on a
system of 24 sites. As it can be observed, all the
techniques predict an energy decreasing with the coupling of
interlayer-dimer via $J_1$. Furthermore, there is an excellent
quantitative agreement between the three methods for small values of $J_{1}$.
On the other hand, triplet gap is shown in the inset of Fig.\ref{fig:results_BO} for the same set of parameters as the ground state
energy. Here we also observe that all the techniques predict a tendency
to a closure of the gap, when $J_1$ is turned on. Our calculations
shows that BO, ED and SE predict the same behavior.

\subsection{Self consistent Schwinger Boson Mean-Field Theory}
\label{sec:sbmft}

As we have seen previously, bond operator and series expansion methods are both suitable to study the interlayer-dimer phase.
In order to investigate the evolution of on-layer phases as a function of inter-layer coupling we apply a  representation of the spin operators in terms of Schwinger bosons \cite{Auerbach}, 
$\vec{\mathbf{S}}_{\alpha}(\vec{r})=\frac{1}{2}\vec{\mathbf{b}}_{\alpha}^{\dag}
(\vec{r})\cdot\vec{\sigma}\cdot\vec{\mathbf{b}}_{\alpha}(\vec{r})$.
Here ${\vec{\bf b}_{\alpha}(\vec{r})^{\dagger }}\!=\!({\bf b}^{\dagger }_{\alpha,\uparrow }(\vec{r}),{\bf b}^{\dagger }_{\alpha,\downarrow }(\vec{r}))$
 is a bosonic spinor corresponding to the site $\alpha$ in the
 unit cell at position $\vec{r}$,  $\vec{\sigma}$ are
Pauli matrices, and the constraint in the number of bosons
$\sum_\sigma \mathbf{b}^{\dag}_{\alpha,\sigma}(\vec{r})\mathbf{b}_{\alpha,\sigma}(\vec{r})\!=\!2S$ has to be satisfied on each site.
In order to perform a mean field decomposition, we define the following $SU(2)$ invariants,
$ \mathbf{A}_{\alpha \beta}(\vec{x},\vec{y})=\frac12 \sum_{\sigma} \sigma
\mathbf{b}_{\alpha,\sigma}(\vec{x})\mathbf{b}_{\beta,-\sigma}(\vec{y})$ and 
$\mathbf{B}_{\alpha \beta}(\vec{x},\vec{y})=\frac12 \sum_{\sigma}
\mathbf{b}^{\dag}_{\alpha,\sigma}(\vec{x})\mathbf{b}_{\beta,-\sigma}(\vec{y})$.
\begin{figure}[t!]
\begin{center}
\includegraphics[width=0.7\columnwidth]{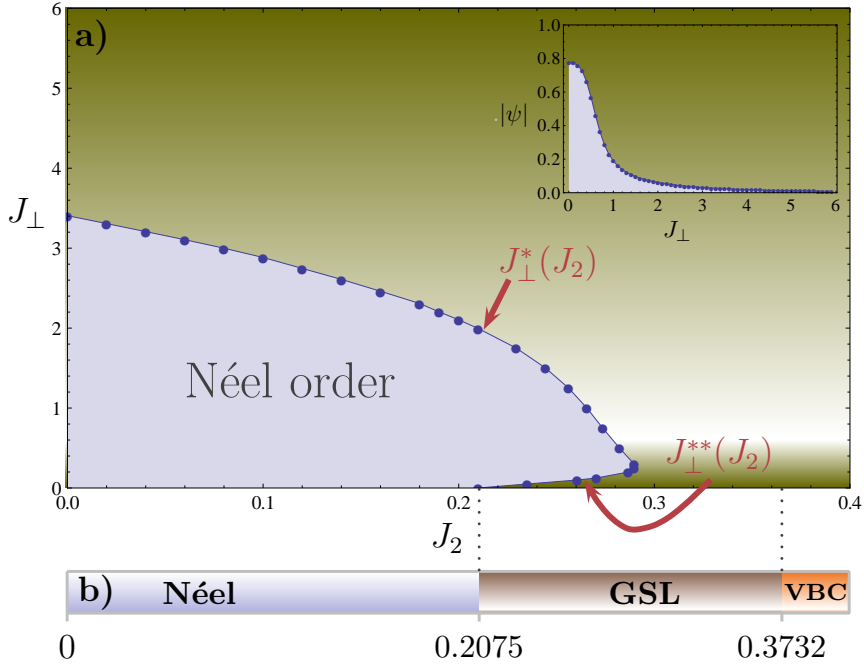}
\caption{ { (Color online) a) Phase diagram in the
$J_{2}-J_{\bot}$ plane obtained with SBMFT. Gray region
corresponds to the N\'eel phase whereas green region corresponds
to magnetically disordered phases.  b) Phase diagram of the single layer case corresponding to
Ref. \cite{Zhang_PRB_2013}. Inset: Z3 order parameter corresponding to the line $J_{2}=0.38$}} \label{fig:Phasediag_Jp-J2}
\end{center}
\end{figure}
This decomposition allow us to treat ferromagnetism and antiferromagnetism on
equal footing and has been successfully used to
describe a number of quantum frustrated
antiferromagnets\cite{Cabra_honeycomb_prb,Cabra_honeycomb_2,Zhang_PRB_2013,Zhang_bilayer,Trumper1,Trumper2,Mezio}.
We perform a Hartree-Fock decoupling where the mean field parameters are the expectation values of
the SU(2) invariants $\mathbf{A}$ and $\mathbf{B}$.
The mean field equations for these parameters $A_{\alpha \beta}$ and $B_{\alpha \beta}$ and the constraints
in the number of bosons must be solved self-consistently (see refs. \cite{Cabra_honeycomb_prb}, \cite{Zhang_PRB_2013} and ,\cite{Zhang_bilayer} for further details). %
To obtain the phase boundary between the magnetically ordered and disordered phases using the self consistent
Schwinger boson mean field theory we study the boson spectrum.
In the gapless region the excitation spectrum is zero at $\vec{k}=\vec{0}$, where the boson condensation occurs,
this is characteristic of the N\'eel ordered phase.
On the other hand, in the gapped region, the absence of Bose condensation indicates that the
ground state is magnetically disordered.

In Fig.~\ref{fig:Phasediag_Jp-J2}-a) we show the phase diagram in the
$J_{2}-J_{\bot}$ plane. For $J_{\bot} \gg
J_2$ one can expect a IVBC ground state adiabatically  connected
with the limit of decoupled dimers, i.e. two singlets per unit cell,
between spins $1,A(1,B)$ and $2,A(2,B)$ (see Fig. \ref{fig:layers}).
%
In the region $0.2075\lesssim J_{2}\lesssim 0.289$ there is a reentrant
effect. In this range, N\'eel phase separates from $J_2$ axis,
leaving a tiny space for a magnetically disordered phase.
In this way, N\'eel phase is here not only limited by some value $J^{*}_{\bot}(J_2)$
 from above, but also by a second value $J^{**}_{\bot}(J_2)$
from below (See Figure \ref{fig:Phasediag_Jp-J2}).

On the other hand, in the range $0.3732 \lesssim J_{2} \lesssim
0.398 $ ($J_{\bot}=0$), there is evidence of the existence of an
on-layer valence bond phase \cite{Zhang_PRB_2013} {(see Figure \ref{fig:Phasediag_Jp-J2}-b)}.
In this phase, SU(2) and translational
symmetries are preserved, but $Z_3$ symmetry is broken.
By turning on $J_{\bot}$ the system moves to the IVBC where the $Z_{3}$
symmetry is recovered. 
In the inset of Figure \ref{fig:Phasediag_Jp-J2}-a,
we depict the $Z_3$ directional symmetry-breaking order parameter
$\rho$ (defined in \cite{Okumura}) vs $J_{\bot}$. 
The behavior of this parameter suggest a $Z_3$ symmetry restoring.
Finally, in the region $0.289 \lesssim J_2 /J_1 \lesssim 0.3732$ the
ground state preserves SU(2), lattice translational and  $Z_3$
symmetries and the spin-spin correlations are short ranged\cite{Zhang_bilayer}. This
agrees with the evidence of a spin liquid phase in the phase diagram
corresponding to $J_{\bot}=0$ \cite{Zhang_PRB_2013,Mezzacapo}.

\section{Conclusions}
In summary, we have studied the phase diagram corresponding to a frustrated
Heisenberg model on the bilayer honeycomb lattice, by means of
bosonic mean field theories, complemented with exact
diagonalization and series expansion, and described the behavior of the quantum phases as the
interlayer coupling is increased.
Using the Schwinger boson description we have determined the region
where the system is N\'eel ordered and the lines where the N\'eel order is destroyed.
We have determined an intermediate region where the
phase diagram shows signatures of a reentrant behavior and we observe that for
values of the interlayer coupling between { ($ 0.289\lesssim
J_{2}/J_{1} \lesssim 0.3732 $)} the N\'eel order is absent at
$J_{\bot}=0$ and the system presents a nonzero spin gap, whereas in
the region { ($0.3732\lesssim J_{2}/J_{1} \lesssim 0.398$)}
 each layer presents a nematic disordered phase\cite{Zhang_PRB_2013}.
In all the range of  $J_{2}$ studied, the system presents signatures of an
interlayer-valence bond crystal (IVBC) phase  for
$J_{\bot}/J_{1}>4$. This phase evolves
adiabatically from the limit of decoupled interlayer-dimers. 
This is corroborated by bond operators self-consistent calculations and series expansions starting explicitly
from the limit of isolated interlayer dimers.

\section*{ACKNOWLEDGMENTS}

C. A. Lamas and M. Arlego are
partially supported by CONICET (PIP 1691) and ANPCyT (PICT 1426 and PICT 2013-0009).



\begin{thebibliography}{}



\bibitem{Cabra_honeycomb_prb}  D. C. Cabra, C. A. Lamas, and H. D. Rosales, Phys. Rev. B 83, 094506
(2011).

\bibitem{Cabra_honeycomb_2}  D. C. Cabra, C. A. Lamas, and H. D. Rosales,  Modern Physics Letters B (MPLB) 25, 891
(2011).

\bibitem{Zhang_PRB_2013} Hao Zhang and C. A. Lamas, Phys. Rev. B 87, 024415
(2013).


\bibitem{Albuquerque}
A.~F.~Albuquerque, D.~Schwandt, B.~Het\'{e}nyi, S.~Capponi,
M.~Mambrini, and A.~M.~L\"auchli, Phys.\ Rev.\ B {\bf 84}, 024406
(2011).

\bibitem{Li_2012_honeyJ1-J2-J3}
P.~H.~Y.~Li, R.~F.~Bishop, D.~J.~J.~Farnell, and C.~E.~Campbell,
Phys.\ Rev.\ B {\bf 86}, 144404 (2012).


\bibitem{QDM_letter} C. A. Lamas, A. Ralko, D. C. Cabra, D. Poilblanc, P. Pujol, Phys. Rev. Lett. {\bf 109}, 016403 (2012)


\bibitem{QDM_lamas} C.A. Lamas, A. Ralko, M. Oshikawa, D. Poilblanc, P. Pujol,  Phys. Rev. B {\bf 87}, 104512 (2013)

\bibitem{Mezzacapo} F.~Mezzacapo and M.~Boninsegni, Phys. Rev. B {\bf 85}, 060402(R)
(2012).

\bibitem{Ganesh_PRL_2013} R. Ganesh, J.~van den Brink, and S. Nishimoto,
\prl{110}{127203}{2013}.


\bibitem{Albuquerque_capponi_2012} Hong-Yu Yang, A F. Albuquerque, S. Capponi, A. M Lauchli and K. P. Schmidt, New J. Phys. {\bf 14} 115027
(2012).


\bibitem{Ganesh_2011} R.\ Ganesh, D.N.\ Sheng, Y.-J.\ Kim and A.\ Paramekanti, Phys.\
Rev.\ B {\bf 83}, 144414 (2011).



\bibitem{Ganesh_QMC} R.\ Ganesh, S.V.\ Isakov, and A.\ Paramekanti, Phys.\
Rev.\ B {\bf 84}, 214412 (2011).

\bibitem{Oitmaa_2012}
J.~Oitmaa and R.~R.~P.~Singh, Phys.\ Rev.\ B {\bf 85}, 014428
(2012).



\bibitem{Zhang_bilayer} Hao Zhang, M. Arlego, C. A. Lamas. Phys. Rev. B {\bf 89}, 024403 (2014).



\bibitem{BiMnO}
O. Smirnova, M. Azuma, N. Kumada, Y. Kusano, M. Matsuda, Y.
Shimakawa, T. Takei, Y. Yonesaki, and N. Kinomura, J. Am. Chem.
Soc., {\bf 131}, 8313 (2009).


\bibitem{ESR} S.~Okubo, F.~Elmasry, W.~Zhang, M.~Fujisawa, T.~Sakurai, H.~Ohta,
M.~Azuma, O.~A.~Sumirnova, and N.~Kumada,, J.\ Phys.: Conf.\ Series
{\bf 200}, 022042 (2010).

\bibitem{Kandpal} H.~C.~Kandpal and J.~van den Brink, Phys. Rev. B {\bf 83}, 140412(R) (2011).

\bibitem{Okumura} S.\ Okumura, H.\ Kawamura, T.\ Okubo, and Y.\ Motome,
J.\ Phys.\ Soc.\ Jpn.\ {\bf 79}, 114705 (2010).


\bibitem{Rastelli} E.\ Rastelli, A.\ Tassi, and L.\ Reatto, Physica B {\bf 97}, 1 (1979).



\bibitem{Fouet} J.\ B.\ Fouet, P.\ Sindzingre, and C.\ Lhuillier, Eur.\ Phys.\ J. B {\bf 20}, 241 (2001).



\bibitem{Einarsson} T.\ Einarsson and H.\ Johannesson, Phys.\ Rev.\ B {\bf 43}, 5867 (1991).



\bibitem{Mattsson_PRB_1994} A. Mattsson, P. Frojdh, and T. Einarsson \prb{49}{3997}{1994}



\bibitem{Oitmaaj1j2j3} J. Oitmaa and R. R. P. Singh,\prb{84}{094424}{2011}

\bibitem{Sachdev} S. Sachdev and R. N. Bath; Phys. Rev: B {\bf 41 } 9323 (1990).



\bibitem{Knetter2000a} C. Knetter and G.S. Uhrig, Eur. Phys. J. B \textbf{13}, 209 (2000).


\bibitem{Arlego2011} M. Arlego and W. Brenig, Phys.\ Rev.\ B {\bf 84}, 134426 (2011).



\bibitem{Auerbach} D.\ P.\ Arovas and A.\ Auerbach,  Phys.\ Rev.\ B {\bf 38}, 316
(1988); A.\ Auerbach and D.\ P.\ Arovas, Phys.\ Rev.\ Lett.\ {\bf
61}, 617 (1988).


\bibitem{Trumper1} H.\ A.\ Ceccato, C.\ J.\ Gazza, and A.\ E.\ Trumper,  Phys.\ Rev.\ B {\bf 47}, 12329 (1993).


\bibitem{Trumper2} A.\ E.\ Trumper, L.\ O.\ Manuel, C.\ J.\ Gazza, and H.\ A.\ Ceccatto,
 Phys.\ Rev.\ Lett.\ {\bf 78}, 2216 (1997).


\bibitem{Mezio} A.~Mezio, C.~N.~Sposetti, L.~O.~Manuel, and A.~E.~Trumper, Europhys. Lett. {\bf 94}, 47001 (2011).




\end{thebibliography}
\end{document}